\documentclass[conference]{IEEEtran}

\usepackage{amsmath}
\usepackage{amssymb}
\usepackage{latexsym}
\usepackage{array,arydshln}
\usepackage{multirow}
\usepackage{graphicx}
\usepackage{float}
\usepackage{bm}
\usepackage{breqn}
\usepackage{ctable}
\usepackage{color}

\newcommand{\otoprule}{\midrule[\heavyrulewidth]}
\newcommand{\bs}[1]{\ensuremath{\boldsymbol{#1}}}
\newcommand{\mBCC}{m}
\newcommand{\mpcc}{m}
\newcommand{\mscc}{m}
\renewcommand{\u}{\bs{u}}
\newcommand{\CU}{C_{\mathrm{U}}}
\newcommand{\CL}{C_{\mathrm{L}}}
\newcommand{\CO}{C_{\mathrm{O}}}
\newcommand{\CI}{C_{\mathrm{I}}}

\newcommand{\ut}[2]{\ensuremath{\bs{u}_{#1,\mathrm{#2}}}}

\newcommand{\fs}[1]{\ensuremath{f_{\mathrm{#1},\mathrm{1}}}}
\newcommand{\fp}[1]{\ensuremath{f_{\mathrm{#1},\mathrm{2}}}}
\newcommand{\fpp}[1]{\ensuremath{f_{\mathrm{#1},\mathrm{3}}}}

\newcommand{\xs}[4]{\ensuremath{p_{\mathrm{#1},\mathrm{#2}}^{(#3,#4)}}}

\IEEEoverridecommandlockouts

\definecolor{red}{rgb}{1,0,0}

\begin{document}

\title{Braided Convolutional Codes -- \\ A Class of Spatially Coupled Turbo-Like Codes}


\author{
\IEEEauthorblockN{Michael Lentmaier$^\dag$, Saeedeh Moloudi$^\dag$, and Alexandre Graell i Amat$^\ddag$}
\IEEEauthorblockA{$\dag$Department of Electrical and Information
  Technology, Lund University, Lund, Sweden \\
  $\ddag$Department of Signals and Systems, Chalmers University of Technology, Gothenburg, Sweden\\
              \{saeedeh.moloudi,michael.lentmaier\}@eit.lth.se, alexandre.graell@chalmers.se}\\
              \thanks{This work was supported in part by the Swedish Research Council (VR) under grant \#621-2013-5477.}\vspace*{-1cm}
}


\maketitle

\begin{abstract} 
In this paper, we investigate the impact of spatial coupling on the
thresholds of turbo-like codes.  Parallel concatenated and serially concatenated convolutional codes as well as braided
convolutional codes (BCCs) are compared by means of an exact density evolution (DE) analysis
for the binary erasure channel (BEC). We propose two extensions of the
original BCC ensemble to improve its threshold and demonstrate that
their BP thresholds approach the maximum-a-posteriori (MAP) threshold
of the uncoupled ensemble.
A comparison of the different ensembles shows that parallel
concatenated ensembles can be outperformed by both serially
concatenated and BCC ensembles, although they have the best BP
thresholds in the uncoupled case.
\end{abstract}

\IEEEpeerreviewmaketitle

\section{Introduction}

It is well known that spatially coupled LDPC codes exhibit a threshold
saturation phenomenon: the threshold of an iterative belief
propagation (BP) decoder, obtained by density evolution (DE), is
improved to that of the optimal maximum-a-posteriori (MAP) decoder
\cite{Kudekar_ThresholdSaturation,LentmaierTransITOct2010}. As a
consequence, it is possible to achieve capacity with simple regular
LDPC codes, which show without spatial coupling a significant gap
between the BP and the MAP threshold. 

The concept of spatial coupling is not limited to LDPC codes. Recently it has been shown that spatial coupling has a similar
effect on the thresholds of turbo-like codes, i.e., concatenated convolutional codes
that can be described by sparse graphical models. Some block-wise
spatially coupled ensembles of  parallel concatenated codes (SC-PCCs)
and serially concatenated codes (SC-SCCs) were introduced in
\cite{MoloudiISTC14}. For the binary erasure channel (BEC) it is possible to
derive exact DE equations for these ensembles from the transfer functions of the component
decoders \cite{Kur03,tenBrinkEXITConv} and perform a threshold
analysis, analogously to \cite{Kudekar_ThresholdSaturation,LentmaierTransITOct2010}.
The numerical results in \cite{MoloudiISTC14} suggest that threshold saturation
occurs if the coupling memory is chosen sufficiently large.
A similar threshold analysis has been performed in \cite{MoloudiISIT14} for braided
convolutional codes (BCCs) \cite{ZhangBCC}, another class of turbo-like codes
which have an inherent spatially coupled structure.

The aim of this paper is to give an overview of spatially coupled turbo-like
codes and compare the thresholds of SC-PCCs, SC-SCCs and BCCs. 
We first present the ensembles considered in \cite{MoloudiISTC14} and
\cite{MoloudiISIT14}. Then we generalize the original BCC ensemble (type-I
BCC) to larger coupling memories and demonstrate that threshold saturation
occurs. Furthermore, inspired by the SC-PCC construction in
\cite{MoloudiISTC14}, we introduce and analyze a new BCC ensemble (type-II BCC) in which
not only the parity symbols but also the information symbols are
coupled over several time instants. The thresholds
are further improved by this construction, i.e., the gap to the MAP
threshold is smaller for a given coupling memory and vanishes as the
memory is increased.

\section{Spatially Coupled Turbo-Like Codes}    
 \label{SC-Turbo}  

In this section, we describe how the concept of spatial coupling can
be applied to turbo-like codes. We consider the parallel and
serially concatenated convolutional codes introduced in \cite{MoloudiISTC14}
as well as the braided convolutional codes from \cite{ZhangBCC} and \cite{MoloudiISIT14}. 
We assume that at each time instant $t$ an information sequence $\bs{u}_t$ is
encoded into a code sequence $\bs{v}_t$, which is transmitted over the
channel. The fundamental idea of spatial coupling is that these
transmitted sequences are interconnected in the encoding process
instead of being processed independently. In order to achieve such an
interconnection, some information symbols and/or code symbols from
previous time instants $t' < t$ serve as inputs of the component
encoders at time $t$. The coupling memory $m$ defines the range of $t'$,
i.e., $t' \in
\{t-m, \dots , t\}$. In what follows we assume that the
encoding starts at $t=1$ and is terminated at $t=L$ in such a way that
$\bs{v}_{L+1}=\bs{0}$.  The value $L$ is called the coupling length.  
Analogously to conventional convolutional
codes, this leads to a rate loss that becomes smaller as $L$
increases.

\subsection{Parallel Concatenated Convolutional Codes}

\begin{figure}[t]
  \centering
    \includegraphics[width=0.75\linewidth]{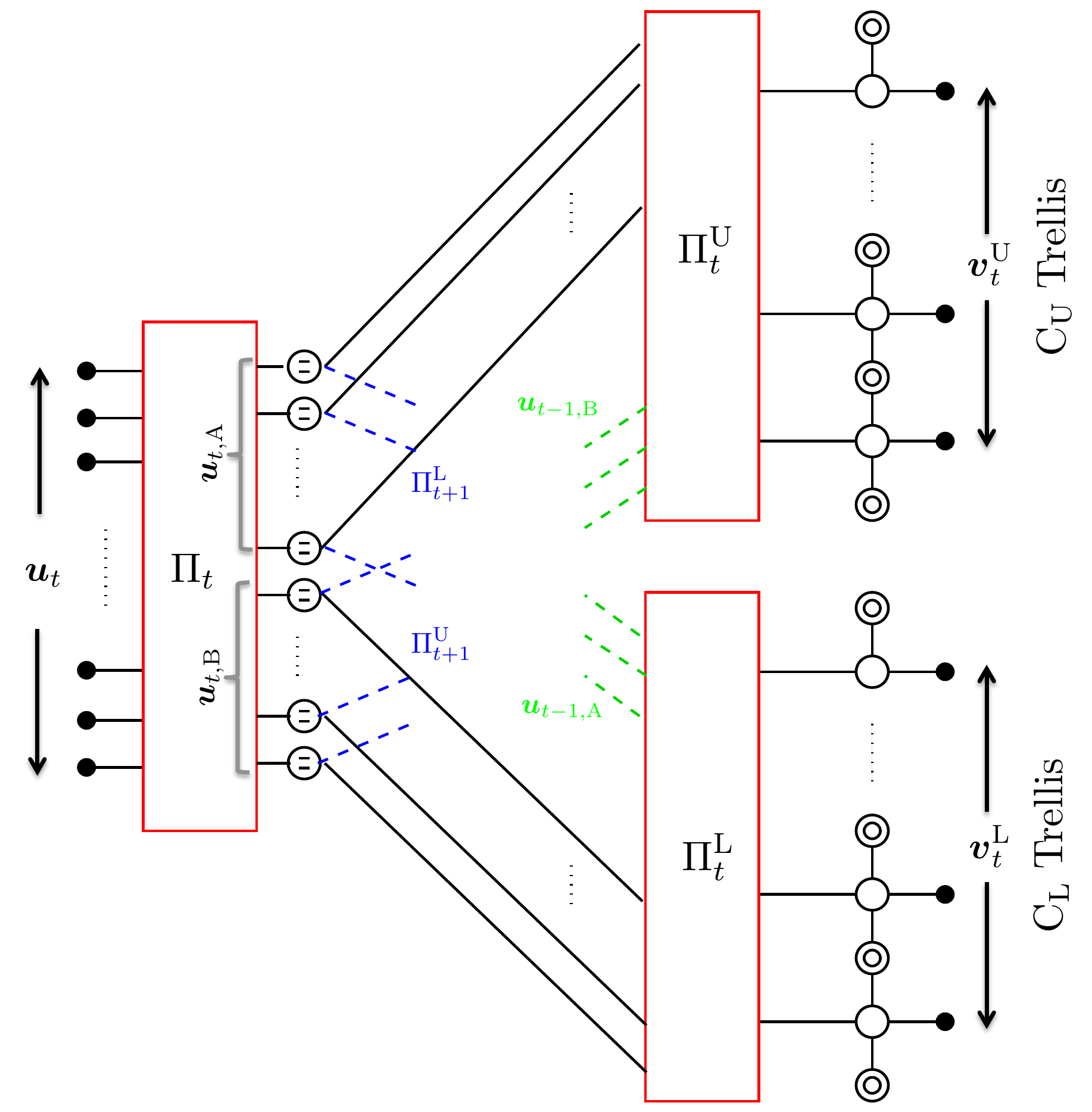}
\caption{Parallel concatenation: factor graph at
  time $t$.}
\label{FactorParal}
\vspace{-2ex}
\end{figure}

Fig.~\ref{FactorParal} shows the factor graph of two parallel
concatenated rate-1/2 convolutional encoders with coupling memory $\mpcc=1$. 
In order to enable a coupled structure the information sequence $\u_t$
is split randomly into two sequences,
$\ut{t}{A}$ and $\ut{t}{B}$. At time $t$ the information sequences
$(\ut{t}{A},\ut{t-1}{B})$ and  $(\ut{t}{B},\ut{t-1}{A})$ are used by
the upper encoder $\CU$ and lower encoder $\CL$ to produce the parity
sequences $\bs{v}^{\text{U}}_t$ and $\bs{v}^{\text{L}}_t$, respectively.
The transmitted code sequence is equal to $\bs{v}_t =
(\bs{u}_t,\bs{v}^{\text{U}}_t ,\bs{v}^{\text{L}}_t )$.
For more details about this code ensemble, including a generalization
to larger coupling memories, we refer the reader to \cite{MoloudiISTC14}.

\subsection{Serial Concatenated Convolutional Codes}
 
\begin{figure}[!t]
  \centering
    \includegraphics[width=0.75\linewidth]{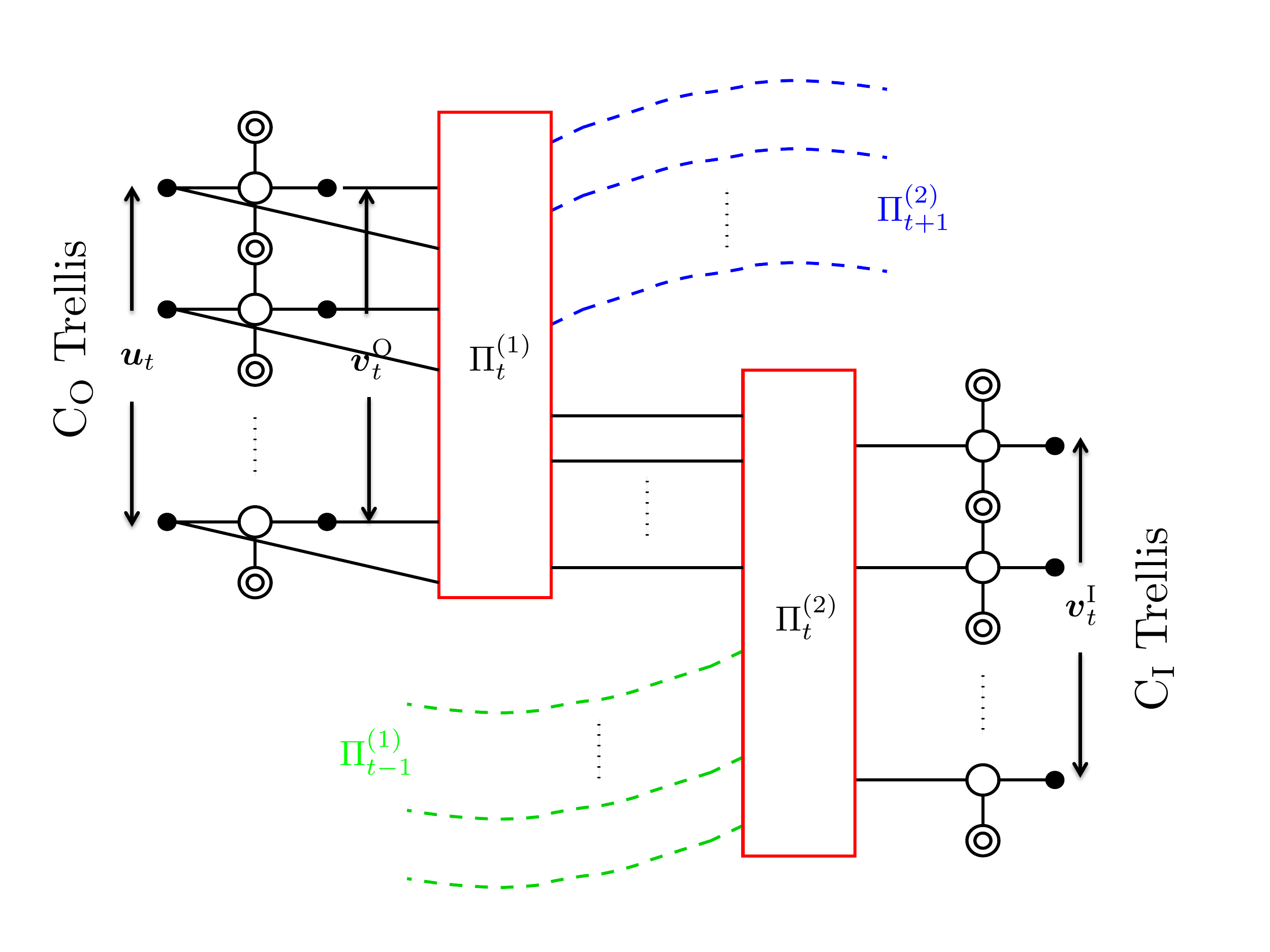}
\caption{Serial concatenation: factor graph at
  time $t$.}
\label{FactorSeri}
\vspace{-2ex}
\end{figure}

The factor graph of two serially concatenated rate-1/2 convolutional
encoders with coupling memory $\mscc=1$ is shown in Fig.~\ref{FactorSeri}. 
In this case the code sequence ${\bs{v}}_t^{\mathrm{O}}$ of the outer
encoder $\CO$ is randomly divided into two parts, $\tilde{\bs{v}}_{t,\text{A}}^{\mathrm{O}}$ and
$\tilde{\bs{v}}_{t,\text{B}}^{\mathrm{O}}$. The input of the inner
encoder $\CI$ at time $t$ is
$(\tilde{\bs{v}}_{t,\text{A}}^{\mathrm{O}},\tilde{\bs{v}}_{t-1,\text{B}}^{\mathrm{O}})$.
The transmitted code sequence $\bs{v}_t$  is equal to the output
of the inner encoder, ${\bs{v}}_t^{\mathrm{I}}$.
For more details about this code ensemble, including a generalization
to larger coupling memories, we refer to \cite{MoloudiISTC14}.

\subsection{Braided Convolutional Codes (Type-I)}
\label{sec:Braidedm1}

\begin{figure}[t]
  \centering
    \includegraphics[width=0.7\linewidth]{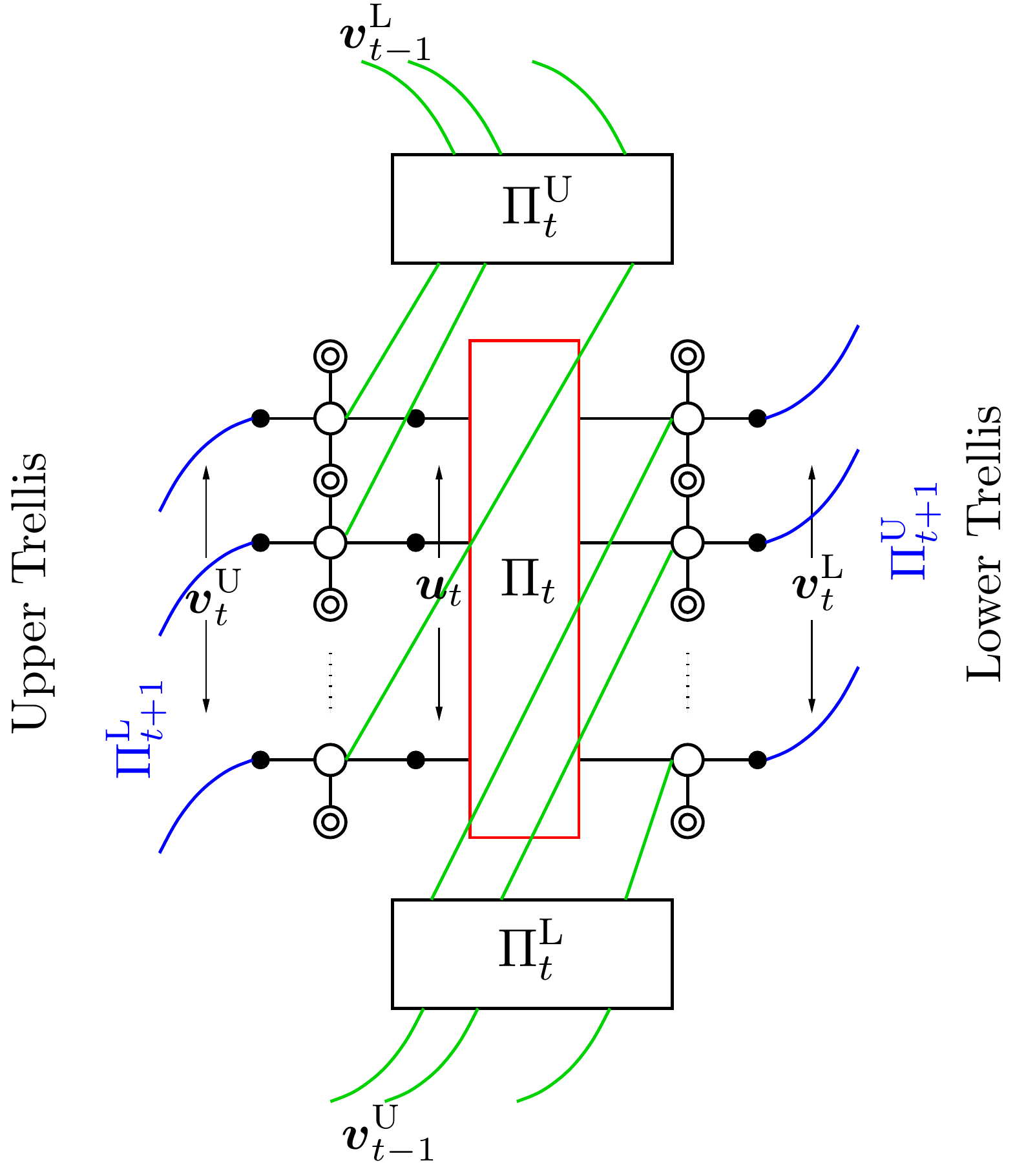}
\caption{Braided convolutional codes (Type-I): factor graph at
  time $t$.}
\label{FactorOriBCC}
\vspace{-2ex}
\end{figure}

Similar to turbo codes, the encoder of a BCC is divided into two
component encoders $\CU$ and $\CL$, with the special feature that the parity sequence of one component
encoder is used as input of the other component encoder. 

\begin{figure*}[!t]
  \centering
    \includegraphics[width=\linewidth]{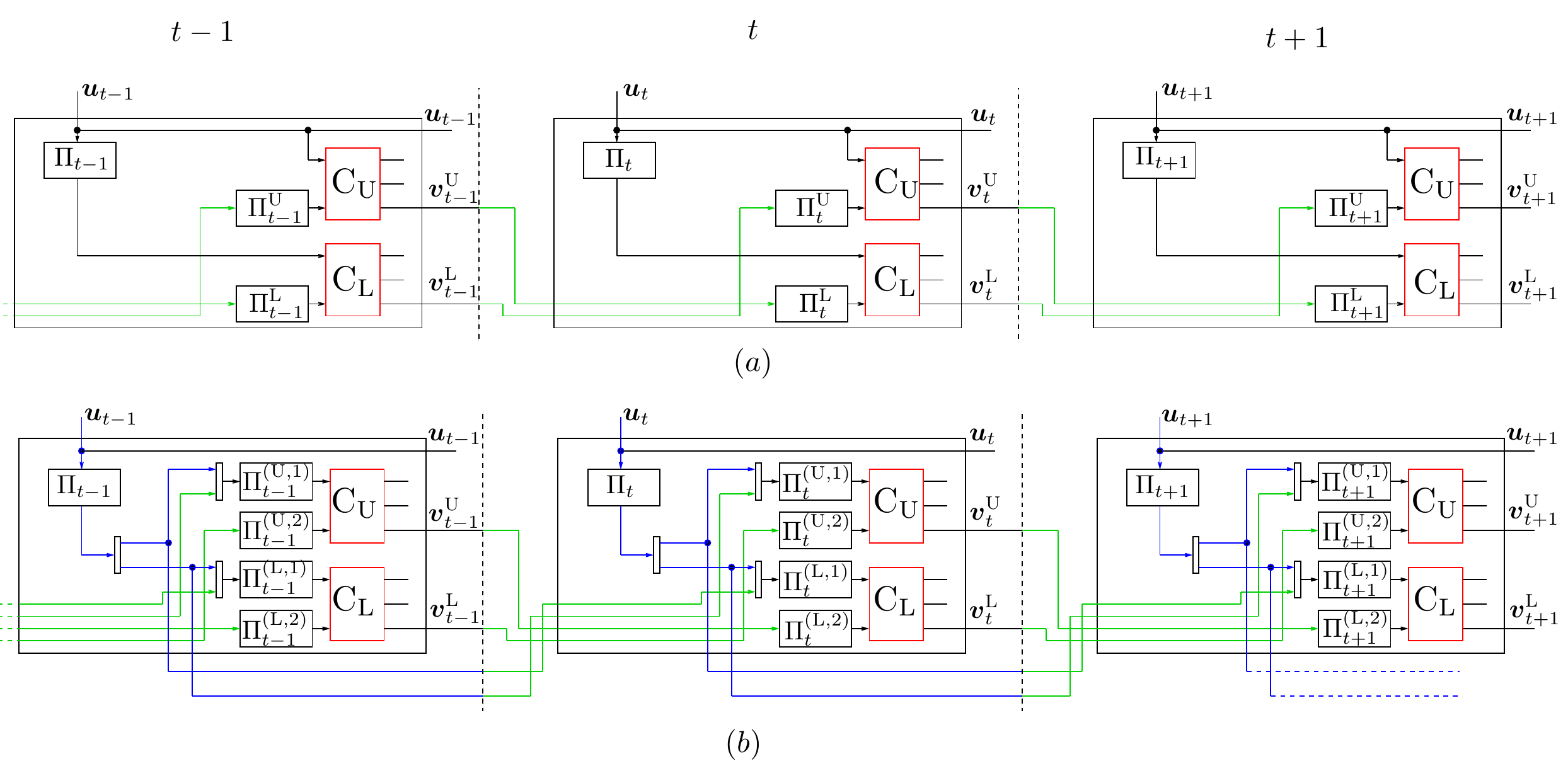}
\caption{Block diagram of the braided convolutional encoder with
  coupling memory $m = 1$. (a) Type-I BCCs (b) Type-II BCCs.}
\label{EncodersM1}
\vspace{-2ex}
\end{figure*}

Fig.~\ref{FactorOriBCC} shows the factor graph of a BCC with
coupling memory $\mBCC=1$, consisting of two
systematic rate-2/3 convolutional encoders, denoted by $\CU$ and $\CL$. 
Their parity sequences  are given by $\bs{v}_{t}^{\text{U}}$ and $\bs{v}_{t}^{\text{L}}$, respectively.
The first input of the upper encoder $\CU$ at time $t$  is 
the information sequence $\bs{u}_t$. The second input is the
previously generated parity sequence $\bs{v}_{t-1}^{\text{L}}$ of $\CL$, after being reordered by a permutation $\Pi_t^{\text{U}}$.
Likewise, the two inputs of the lower encoder $\CL$ are $\bs{u}_t$ and 
$\bs{v}_{t-1}^{\text{U}}$, which are both properly reordered by the permutations $\Pi_t$
and $\Pi_{t}^{\text{L}}$, respectively.  
The transmitted code sequence is
$\bs{v}_{t}=(\bs{u}_{t},\bs{v}_{t}^{\text{U}},\bs{v}_{t}^{\text{L}})$.
An encoder block diagram of the resulting BCC with overall rate $R=1/3$ is
illustrated in Fig.~\ref{EncodersM1}(a).

The above described ensemble is equivalent to the original block-wise BCC ensemble
introduced in \cite{ZhangBCC} and analyzed in \cite{MoloudiISIT14}. Throughout this
paper we will refer to these codes as type-I BCCs. 
                                                                                                                                                                                                      
\section{Type-I Braided Convolutional Codes: Generalization to Larger Coupling Memories}
\label{Braided-type1}

In this section, we generalize the type-I BCCs described in Section~\ref{sec:Braidedm1} to coupling memory
$\mBCC>1$. 

\begin{figure}[!t]
  \centering
    \includegraphics[width=\linewidth]{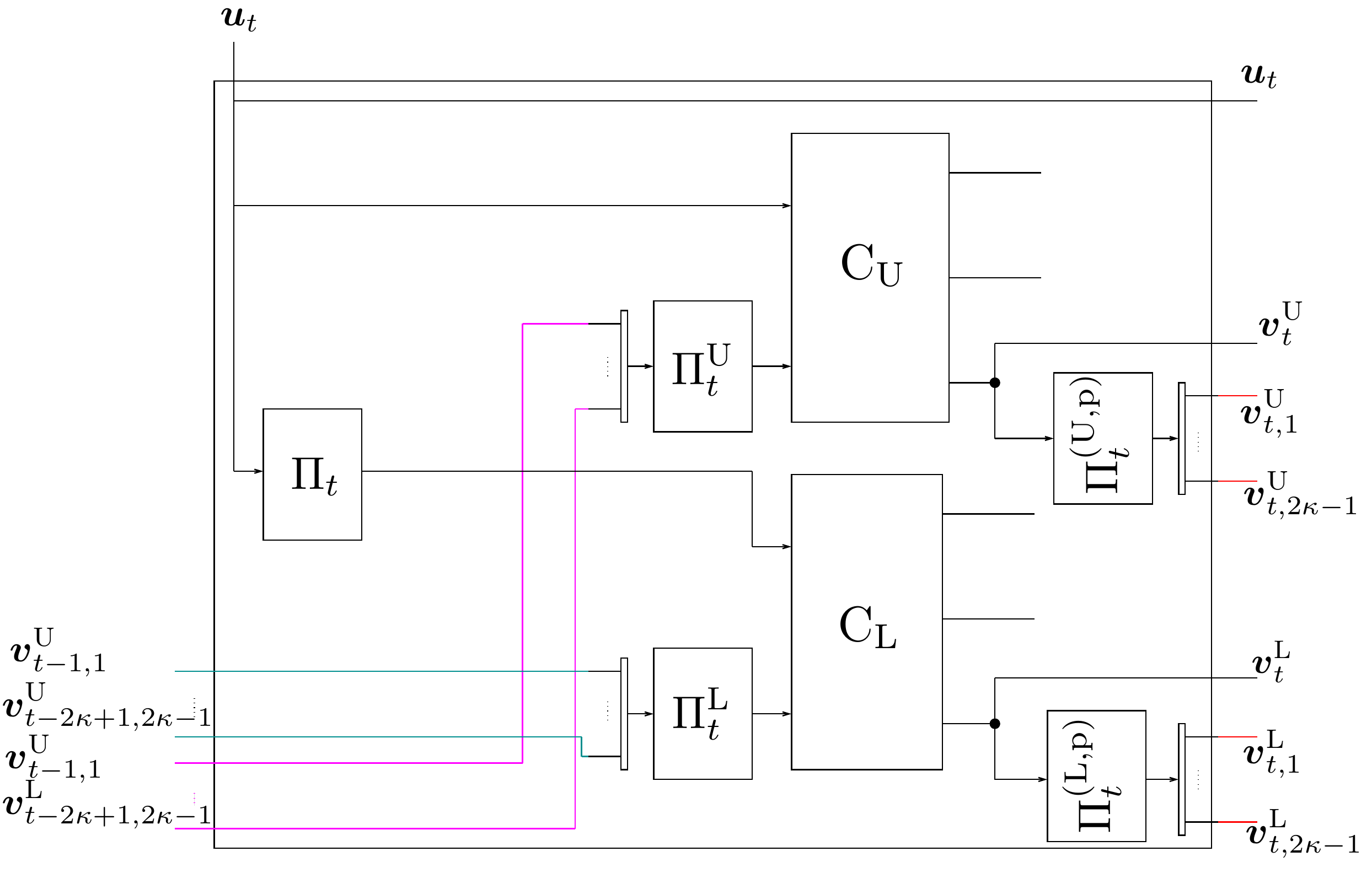}
\caption{Block diagram of the encoder of type-I BCCs with $\mBCC=2\kappa-1$.}
\label{EncodersG1}
\vspace{-2ex}
\end{figure}

Fig.~\ref{EncodersG1} shows the encoder of type-I BCCs with coupling memory $\mBCC=2\kappa-1$. At time $t$, the coded sequence of $\CU$,
$\bs{v}_{t}^{\text{U}}$, is randomly divided into $\mBCC$ parts by the
use of permutation $\Pi_{t}^{(\text{U,p})}$ followed by a
demultiplexer. These $\mBCC$ sequences are shown in the right side of Fig.~\ref{EncodersG1} and denoted by
$\bs{v}_{t,j}^{\text{U}}, j=1,\dots,2\kappa-1$. At time $t$, the sequences $\bs{v}_{t-j,j}^{\text{U}}$ are multiplexed and reordered by the permutation $\Pi_{t}^{\text{L}}$ to create a new sequence, which is used as a second input of $\CL$. Likewise, the coded sequence of the lower encoder at time $t$ is split into $\mBCC$ parts,
$\bs{v}_{t,j}^{\text{L}}$, $j=1,\dots,\mBCC$. The sequences $\bs{v}_{t-j,j}^{\text{L}}$ are then merged and reordered by permutation $\Pi_{t}^{\text{U}}$ to
create the second input of $\CU$. On the other hand, for type-I BCCs, the first
input of $\CU$ and $\CL$ are the information bits and reordered
information bits, respectively.

\subsection{Density Evolution Analysis} \label{generalDE}

The extrinsic erasure probability of the $k$th output of $\CU$
is denoted by $\ensuremath{p_{\mathrm{U},\mathrm{k}}}$, $k=1,2,3$. These
erasure probabilities $\ensuremath{p_{\mathrm{U},\mathrm{k}}}$, at iteration $i$ and time $t$, can be written as
\begin{align}
\label{eq:UpperUpdate1}
\xs{U}{1}{i}{t}&=\fs{U}\left(
q_{\text{L},1}^{(i-1)},q_{\text{L},2}^{(i-1)},q_{\text{L},3}^{(i-1)}\right),\\
\label{eq:UpperUpdate2}
\xs{U}{2}{i}{t}&=\fp{U}\left(
q_{\text{L},1}^{(i-1)},q_{\text{L},2}^{(i-1)},q_{\text{L},3}^{(i-1)}\right),\\
\label{eq:UpperUpdate3}
\xs{U}{3}{i}{t}&=\fpp{U}\left(
q_{\text{L},1}^{(i-1)},q_{\text{L},2}^{(i-1)},q_{\text{L},3}^{(i-1)}\right).
\end{align}
where $\ensuremath{f_{\mathrm{U},\mathrm{k}}}$ is the decoder
transfer function of $\CU$ for the $k$th output of the decoder, and
$\ensuremath{q_{\mathrm{L},\mathrm{k}}}$ 
are the input erasure probabilities. In~\cite{MoloudiISIT14} we described the method to derive the exact
expression for the transfer functions
$\ensuremath{f_{\mathrm{U},\mathrm{k}}}$ and compute DE equations for the case $\mBCC=1$. Here, we generalize the equations to coupling memory $\mBCC$.

The input erasure probabilities
$q_{\text{L},1}^{(i-1)},q_{\text{L},2}^{(i-1)}$ and
$q_{\text{L},3}^{(i-1)}$ in (\ref{eq:UpperUpdate1}),
(\ref{eq:UpperUpdate2}) and (\ref{eq:UpperUpdate3})
are as follow,
\begin{equation}
\label{eq:UpperUpdate3general1}
q_{\text{L},1}^{(i-1)}=\epsilon \cdot \xs{L}{1}{i-1}{t}  ,
\end{equation}
\begin{equation}
\label{eq:UpperUpdate3general2}
q_{\text{L},2}^{(i-1)}=\epsilon \cdot\frac{\sum_{j=1}^{\mBCC}\xs{L}{3}{i-1}{t-j}}{\mBCC} \ ,
\end{equation}
\begin{equation}
\label{eq:UpperUpdate3general3}
q_{\text{L},3}^{(i-1)}=\epsilon \cdot\frac{\sum_{j=1}^{\mBCC}\xs{L}{2}{i-1}{t+j}}{\mBCC} \ ,
\end{equation}
where $\xs{L}{k}{i}{t}$ is the extrinsic erasure probability of the $k$th output of $\CL$ at iteration $i$ and time $t$. Note that $\xs{L}{k}{i}{t}$  is equal to zero
 for $t<0$ and $t>L$ and it is 1 for
 $i=0$. The equations for the
lower decoder are obtained by simply interchanging the indices $\text{U}$ and $\text{L}$
with each other in (\ref{eq:UpperUpdate1})--(\ref{eq:UpperUpdate3general3})

Finally, the a-posteriori erasure probability on information bits at
iteration $i$ and time $t$ is
 \begin{equation}
\label{eq:appPCC}
p^{(i,t)}_{\rm a}=\epsilon\cdot 
\xs{U}{1}{i}{t}\cdot\xs{L}{1}{i}{t}.
\end{equation}

\section{Type-II Braided Convolutional Codes: Coupling of Information Symbols}
\label{Braided-type2}

For simplicity, we first describe the coupling with coupling
memory $m=1$ and we then consider the generalization to 
$m>1$.
 
\subsection{Coupling of Information Symbols}

\begin{figure}[!t]
 \centering
 \includegraphics[width=.75\linewidth]{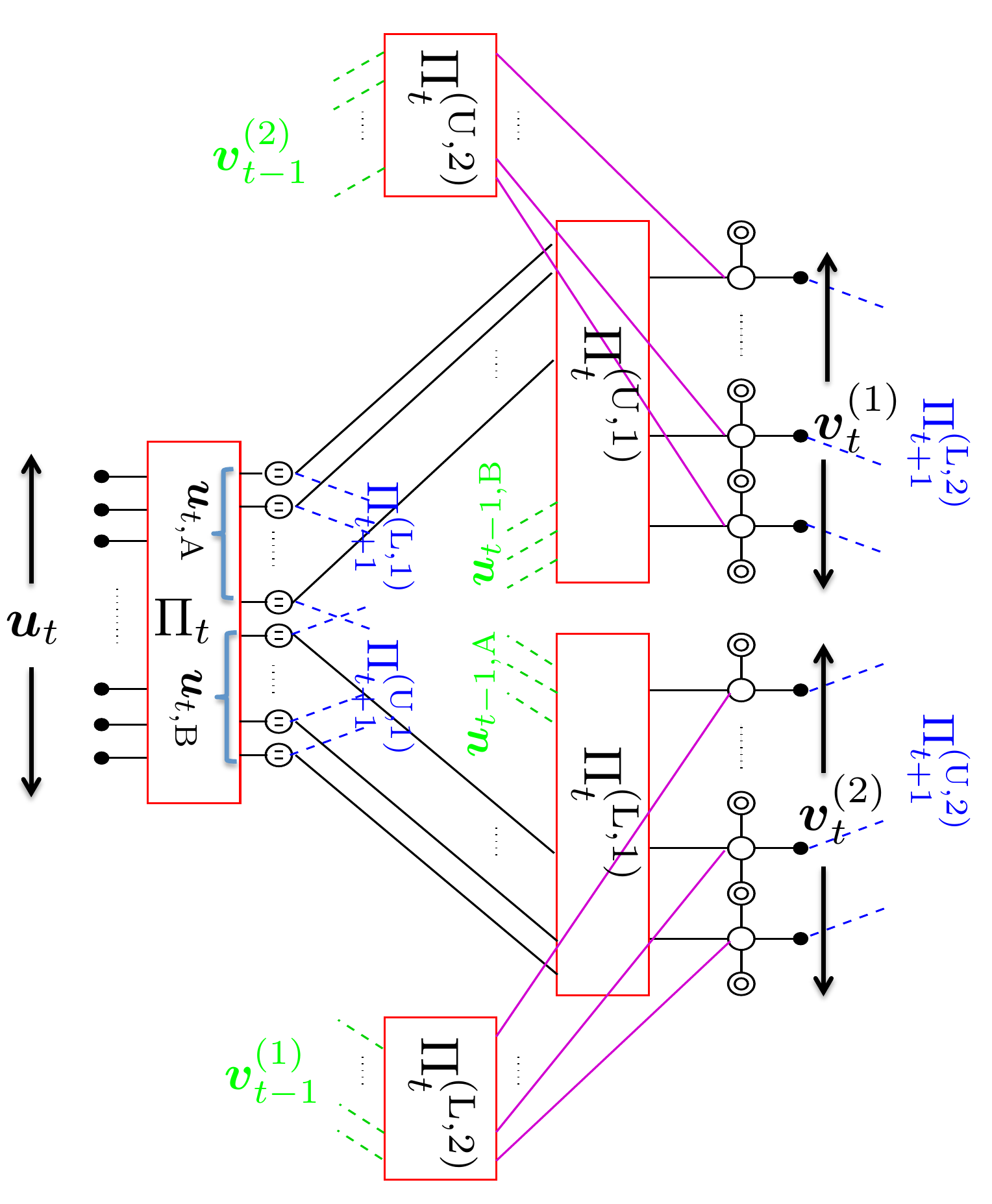}
\caption{Braided convolutional codes (Type-II): factor graph at
  time $t$.}
\label{factorBCC}
\vspace{-2ex}
\end{figure}

The factor graph of type-II BCCs with
$\mBCC=1$ is shown in Fig.~\ref{factorBCC}.
Consider a chain of $L$ encoders between time instants $t=1$
and $t=L$. According to Fig.~\ref{factorBCC}, divide randomly the information sequence $\u_t$ at time $t$ into two
parts, $\ut{t}{A}$ and $\ut{t}{B}$. The
information sequence $(\ut{t}{A},\ut{t-1}{B})$, properly reordered by
a permutation $\Pi^{(\text{U},1)}_t$ is then used as the first input of encoder
$\CU$. Likewise, the sequence at the first input of encoder $\CL$ is
$(\ut{t}{B},\ut{t-1}{A})$, 
properly reordered by the permutation $\Pi^{(\text{L},1)}_t$. The second
inputs of $\CU$ and $\CL$ are the same as that of the original braided codes, $\bs{v}_{t-1}^{\text{L}}$ and
$\bs{v}_{t-1}^{\text{U}}$, reordered by 
permutations $\Pi^{(\text{U},2)}_t$ and $\Pi^{(\text{L},2)}_t$, respectively. This coupling method is also
illustrated in Fig.~\ref{EncodersM1}(b). In the figure, blue lines
represent bits from the current time instant and green lines represent
bits from the previous time instant.


\subsection{Generalization to Larger Coupling Memories}


For simplicity, we limit ourselves to odd values of $\mBCC$, $\mBCC=2\kappa-1$ for some $\kappa\ge 1$. 

\begin{figure}[!t]
  \centering
    \includegraphics[width=\linewidth]{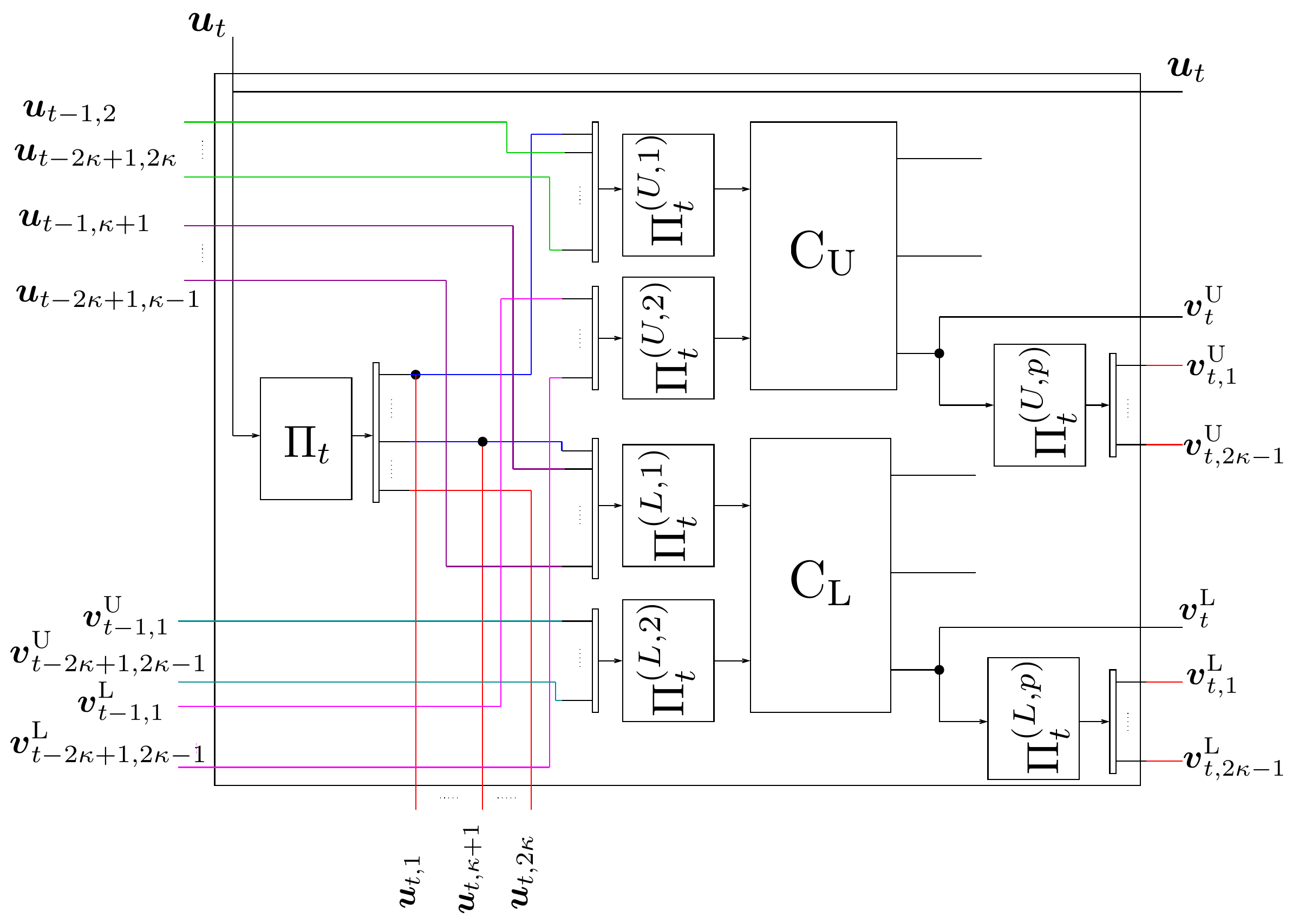}
\caption{Block diagram of the encoder of type-II BCCs with $\mBCC=2\kappa-1$.}
\label{EncodersG2}
\vspace{-2ex}
\end{figure}

Fig.~\ref{EncodersG2} shows the general form of type-II BCCs. In the general case, the information
sequences $\bs{u}_{t}, \bs{u}_{t-1}, \dots, \bs{u}_{t-2\kappa+1}$ and
parity sequences $\bs{v}_{t-1}^{\text{L}}, \bs{v}_{t-2}^{\text{L}}, \dots, \bs{v}_{t-2\kappa+1}^{\text{L}}$ from
different time instants contribute to the inputs of $\CU$ at time
$t$.  Likewise, sequences  $\bs{u}_{t}, \bs{u}_{t-1}, \dots, \bs{u}_{t-2\kappa+1}$ and
$\bs{v}_{t-1}^{\text{U}}, \bs{v}_{t-2}^{\text{U}}, \dots,
\bs{v}_{t-2\kappa+1}^{\text{U}}$ contribute to the inputs of $\CL$. This is achieved by randomly dividing the information sequence
$\bs{u}_t$ and the parity sequences of $\CU$ and $\CL$,
$\bs{v}^{\text{U}}_t$ and $\bs{v}^{\text{L}}_t$, into the sequences $\bs{u}_{t,j}$,
$j=1,\dots,2\kappa$, and $\bs{v}^{\text{U}}_{t,l},
\bs{v}^{\text{L}}_{t,l}$, $l=1,\dots, 2\kappa-1$, with permutations
$\Pi_t$, $\Pi_t^{\text{(U,p)}}$ and $\Pi_t^{\text(L,p)}$, respectively. The first input of $\CU$
at time $t$ is the sequence obtained by multiplexing and
reordering the sequences $\bs{u}_{t-j+1,j}$ by the permutation $\Pi_{t}^{\text{(U,1)}}$. The second input of $\CU$  is the sequence obtained by multiplexing and reordering (through permutation $\Pi_{t}^{\text{(U,2)}}$) the sequences $\bs{v}_{t-l,l}^{\text{L}}$.

The lower encoder
$\CL$ receives as first input the complementary set of information sequences
$\bs{u}_{t-j+1,j'}$, $j=1,\dots,2\kappa$, $j'=(\kappa+j-1 \mod
2\kappa)+1$ in a symmetric fashion, multiplexed and reordered by
$\Pi_{t}^{\text{(L,1)}}$. The second
input of $\CL$ is the sequence obtained by multiplexing and reordering (through $\Pi_{t}^{\text{(U,2)}}$) the sequences $\bs{v}_{t-l,l}^{\text{U}}$. It follows that the information sequence $\bs{u}_{t,j}$ is used by
$\CU$ at time $t+j-1$ and by $\CL$ at time $t+(\kappa+j-1 \mod
2\kappa)$. The encoder in Fig.~\ref{factorBCC} and Fig.~\ref{EncodersM1}(b) corresponds to the special case
$\kappa=1$. 

\subsection{Density Evolution Analysis}

Considering type-II BCCs with $\mBCC=1$,
the input erasure probabilities at the input of the upper decoder (to be used in
(\ref{eq:UpperUpdate1}),
(\ref{eq:UpperUpdate2}) and (\ref{eq:UpperUpdate3}))
are

\begin{equation}
\label{eq:UpperUpdateCouInfM11}
q_{\text{L},1}^{(i-1)}=\epsilon \cdot\frac{\xs{L}{1}{i-1}{t-1}+\xs{L}{1}{i-1}{t+1}}{2} \ ,
\end{equation}

\begin{equation}
\label{eq:UpperUpdateCouInfM12}
q_{\text{L},2}^{(i-1)}=\epsilon \cdot\xs{L}{3}{i-1}{t-1}\ ,
\end{equation}

\begin{equation}
\label{eq:UpperUpdateCouInfM13}
q_{\text{L},3}^{(i-1)}=\epsilon \cdot\xs{L}{2}{i-1}{t+1}\ .
\end{equation}

We remark that these equations are identical to the ones in \cite{MoloudiISIT14} for type-I BCCs except for (\ref{eq:UpperUpdateCouInfM11}). 
 
Finally, the a-posteriori erasure probability on information bits at
time $t$ and iteration $i$ is
\begin{equation}
\label{eq:appPCCinf}
p^{(i,t)}_{\rm a}=\epsilon\cdot \frac{\xs{U}{1}{i}{t}\xs{L}{1}{i}{t+1}+\xs{U}{1}{i}{t+1}\xs{L}{1}{i}{t}}{2}.
\end{equation}

It is possible to generalize (\ref{eq:UpperUpdateCouInfM11})--(\ref{eq:appPCCinf}) to $\mBCC=2\kappa-1$. We obtain
\begin{equation}
\label{eq:UpperUpdate3generalb}
q_{\text{L},1}^{(i-1)}=\epsilon \cdot\frac{\xs{L}{1}{i-1}{t-\kappa}+\xs{L}{1}{i-1}{t+\kappa}}{2} \ ,
\end{equation}
while $q_{\text{L},2}^{(i-1)}$ and $q_{\text{L},3}^{(i-1)}$ are the same as
(\ref{eq:UpperUpdate3general2}) and (\ref{eq:UpperUpdate3general3}) for type-I BCCs (see Section~ \ref{generalDE}) by setting $\mBCC=2\kappa-1$.

For the lower decoder we can obtain similar expressions by interchanging
the indices $\text{U}$ and $\text{L}$ in (\ref{eq:UpperUpdateCouInfM11})--(\ref{eq:UpperUpdateCouInfM13}) and (\ref{eq:UpperUpdate3generalb}).

The a posteriori erasure probability on the information bits at time $t$ and iteration $i$ (\ref{eq:appPCC}) becomes
 \[
p^{(i,t)}_{\rm a}=\epsilon\cdot \frac{\sum_{j=1}^{2\kappa} \xs{U}{1}{i}{t+j-1}\xs{L}{1}{i}{t+(\kappa+j-1) \mathrm{mod} 2\kappa}}{2\kappa} \ .
\]

\section{Results and Discussion}
\label{Result}

In this section, we present the BP thresholds for the two BCC
ensembles considered in Section~\ref{Braided-type1} and
~\ref{Braided-type2} and compare them to SC-PCCs and SC-SCCs
\cite{MoloudiISTC14}. In particular, we consider BCCs with the two
identical 4-state rate-2/3 component
encoders used in \cite{ZhangBCC}, with generator matrix
\begin{equation*}\label{eqG}
\bs{G} (D)=\left( \begin{array}{ccc}1&0&\frac{1}{1+D+D^2}\\0&1&\frac{1+D^2}{1+D+D^2}\end{array}\right)
= \left( \begin{array}{ccc}1&0&1/7\\0&1&5/7\end{array}\right) \ .
\end{equation*}
Likewise, for SC-PCCs and SC-SCCs we consider two identical 4-state
rate-1/2 component encoders with generator matrix $\bs{G}=(1, \, 5/7)$
(in octal notation). In order to compare the SC-SCCs with the other
ensembles we increase their rate to $R=1/3$ using the random puncturing procedure described in \cite{MoloudiISTC14}.
All presented thresholds correspond to the stationary case  $L \rightarrow \infty$, which lower bounds the thresholds for finite $L$.  For small $L$ the threshold can be considerably larger but at the expense of a higher rate loss.

In Table~\ref{Tab:BCC} we give the BP threshold $\epsilon_{\mathrm{SC}}$ for BCCs of type-I and
type-II with different coupling memory $m$. We also report in the table
the BP threshold $\epsilon_{\mathrm{BP}}$ and the MAP threshold $\epsilon_{\mathrm{MAP}}$ of the
uncoupled ensembles.  The MAP threshold was computed applying the area
theorem \cite{Measson2009}.  Note that the value for type-I with $m=1$ corresponds
to the original ensemble considered in \cite{MoloudiISIT14}. It can be
observed that $\epsilon_{\mathrm{SC}}$ can be improved by increasing $m$ and it is
expected that threshold saturation occurs. We can also see that the
thresholds are better for type-II BCCs and are very close to the MAP
threshold already for small coupling memory $m$.

The corresponding thresholds for SC-PCCs and punctured
SC-SCCs are presented in Table~\ref{Tab:PCC}. Note that the puncturing rates are taken from
  \cite{MoloudiISTC14} and are not necessarily the best possible
  choices. An optimization of these coefficients could lead to
  improvements regarding either the BP or the MAP threshold. We can observe that SC-PCCs have the
best BP threshold in the uncoupled case but have a MAP threshold that
is relatively poor. With spatial coupling a saturation to this MAP threshold
occurs already for small coupling memory $m$. BCCs, on the other hand,
have the worst BP threshold without coupling but a better MAP
threshold than both SC-PCCs and SC-SCCs. In terms of the  thresholds
$\epsilon_{\mathrm{SC}}$ with spatial coupling we can see that BCCs of type-I and
type-II outperform the other ensembles for all coupling memories. This
confirms that BCCs are a powerful class of codes. Note that
asymptotically they are also known to have superior distance properties (their
minimum distance grows linearly with the block length \cite{ZhangBCC}).

\begin{table}[t]
\caption{Thresholds for BCCs ($R=1/3$)}
\begin{center}
\scalebox{0.9}{\begin{tabular}{lcccccc}
\toprule
& $\epsilon_{\text{BP}}$ & $\epsilon_{\text{MAP}}$&
\multicolumn{4}{c}{$\epsilon_{\text{SC}}$} \\[0.5mm]
& & & $m=1$ &$m=3$ & $m=5$ & $m=7$ \\
\otoprule
Type-I  & 0.55414 & 0.66539 & 0.66094 & 0.66447 & 0.66506 & 0.66524\\[0.5mm]
Type-II & 0.55414 & 0.66539 & 0.66534 & 0.66538 & 0.66539 & 0.66539\\
\bottomrule
\end{tabular}} \end{center}
\label{Tab:BCC} 
\vspace{-3ex}
\end{table}

\begin{table}[t]
\caption{Thresholds for SC-PCCs and punctured SC-SCCs ($R=1/3$)}

\begin{center}
\scalebox{0.9}{\begin{tabular}{lccccc}
\toprule
& $\epsilon_{\text{BP}}$ & $\epsilon_{\text{MAP}}$&
\multicolumn{3}{c}{$\epsilon_{\text{SC}}$} \\[0.5mm]
& & & $m=1$ &$m=3$ & $m=5$  \\
\otoprule
SC-PCCs & 0.64282 & 0.65538 & 0.65538 & 0.65538 & 0.65538 \\[0.5mm]
SC-SCCs & 0.61184 & 0.66154 & 0.65190 & 0.66140 & 0.66153 \\
\bottomrule
\end{tabular}} \end{center}
\label{Tab:PCC} 
\vspace{-3ex}
\end{table}

\newpage 

\section{ Conclusions}
\label{Conclud}

In this paper we investigated the impact
of spatial coupling on the BP  decoding threshold of turbo-like codes. 
BCCs are a powerful class of turbo-like  codes
which are closely related to LDPC codes and product codes. 
We introduced two novel BCC ensembles (type-I and type-II) that generalize the original
ensembles in \cite{ZhangBCC}. For these ensembles we derived exact density evolution
recursions for the BEC, which allowed us to numerically evaluate the
thresholds for different coupling memories $m$ and compare them to the
thresholds of parallel and serially concatenated convolutional codes.
For all three classes of codes it can be observed that the BP
threshold $\epsilon_{\text{SC}}$ improves with
increasing $m$, and we assume that a saturation to the MAP threshold  $\epsilon_{\text{MAP}}$  
occurs. The best thresholds are obtained for the novel BCC ensemble of
type-II. Interestingly, the gap between the BP threshold $\epsilon_{\text{BP}}$ and the MAP
threshold $\epsilon_{\text{MAP}}$  is largest for the BCCs, which perform worst in the
uncoupled case.

The considered examples of turbo-like codes demonstrate that the concept of spatial
coupling opens some new degrees of freedom in the design of
codes on graphs: instead of optimizing the component encoder characteristics for
BP decoding it is possible to optimize the MAP decoding threshold and
rely on the threshold saturation effect of spatial coupling. 
Powerful code ensembles with strong distance properties can then
perform close to capacity with low-complexity iterative decoding.







\begin{thebibliography}{1}

\bibitem{Kudekar_ThresholdSaturation}
S.~Kudekar, T.J. Richardson, and R.L. Urbanke,
\newblock ``Threshold saturation via spatial coupling: {W}hy convolutional
  {LDPC} ensembles perform so well over the {BEC},''
\newblock {\em {IEEE} Trans. Inf. Theory}, vol. 57, no. 2, pp. 803 --834, Feb.
  2011.

\bibitem{LentmaierTransITOct2010}
{M.~Lentmaier}, {A.~Sridharan}, {D.J.~Costello, Jr.}, and {K.Sh.~Zigangirov},
\newblock ``Iterative decoding threshold analysis for {LDPC} convolutional
  codes,''
\newblock {\em IEEE Trans.~Inf.~Theory}, vol. 56, no. 10, pp. 5274--5289, Oct.
  2010.

\bibitem{MoloudiISTC14}
S.~Moloudi, M.~Lentmaier, and A.~Graell~i Amat,
\newblock ``Spatially coupled turbo codes,'' submitted to {\em 8’th
  International Symposium on Turbo Codes \& Iterative Information Processing},
  2014.

\bibitem{Kur03}
B.M. Kurkoski, P.H. Siegel, and J.K. Wolf,
\newblock ``Exact probability of erasure and a decoding algorithm for
  convolutional codes on the binary erasure channel,''
\newblock in {\em Proc.~IEEE Global Telecommunications Conference, 2003.
  GLOBECOM '03.}, Dec. 2003, vol.~3.

\bibitem{tenBrinkEXITConv}
J.~Shi and S.~ten Brink,
\newblock ``Exact {EXIT} functions for convolutional codes over the binary
  erasure channel,''
\newblock in {\em Proceedings of the 44th Allerton Conference on Communication,
  Control, and Computing}, Monticello, IL, USA, 2006.

\bibitem{MoloudiISIT14}
S.~Moloudi and M.~Lentmaier,
\newblock ``Density evolution analysis of braided convolutional codes on the
  erasure channel,''
\newblock in {\em Proc. IEEE International Symposium on Information Theory},
  Honolulu, HI, USA, July 2014.

\bibitem{ZhangBCC}
{W. Zhang}, {M. Lentmaier}, {K.Sh. Zigangirov}, and {D.J. Costello, Jr.},
\newblock ``Braided convolutional codes: a new class of turbo-like codes,''
\newblock {\em {IEEE} Trans. Inf. Theory}, vol. 56, no. 1, pp. 316--331, Jan.
  2010.

\bibitem{Measson2009}
C.~Measson, A.~Montanari, T.J. Richardson, and R.~Urbanke,
\newblock ``The generalized area theorem and some of its consequences,''
\newblock {\em IEEE Trans.~Inf.~Theory}, vol. 55, no. 11, pp. 4793--4821, Nov.
  2009.

\end{thebibliography}
\end{document}